\RequirePackage{lineno}
\documentclass[twocolumn,preprintnumbers,amsmath,amssymb]{revtex4}
\usepackage{graphicx}
\usepackage{dcolumn}
\usepackage{bm}
\usepackage{color}
\usepackage[final]{pdfpages}
\begin{document}
\title{Metallicity without quasi-particles in room-temperature strontium titanate}
\author{Xiao Lin$^{1}$, Carl Willem Rischau$^{1}$, Lisa Buchauer$^{1}$, Alexandre Jaoui$^{1}$, Beno\^{\i}t Fauqu\'e$^{1,2}$  and Kamran Behnia$^{1}$}
\affiliation{(1) Laboratoire Physique et Etude de Mat\'{e}riaux (CNRS-UPMC), ESPCI Paris, PSL Research University, 75005 Paris, France\\
(2) JEIP,  USR 3573 CNRS, Coll\`{e}ge de France, PSL Research University,  75005 Paris, France\\}
\date{April 28, 2017}

\begin{abstract}
Cooling oxygen-deficient strontium titanate to liquid-helium temperature leads to a decrease in its  electrical resistivity by several orders of magnitude. The temperature dependence of resistivity follows a rough T$^{3}$ behavior before becoming T$^{2}$ in the low-temperature limit, as expected in a Fermi liquid. Here, we show that the roughly cubic resistivity above 100K corresponds to a regime where the quasi-particle mean-free-path is shorter than the electron wave-length and the interatomic distance. These  criteria define the Mott-Ioffe-Regel limit. Exceeding this limit is the hallmark of strange metallicity, which occurs in strontium titanate well below room temperature, in contrast to other perovskytes. We argue that the T$^{3}$-resistivity cannot be accounted for by electron-phonon scattering \`{a} la Bloch-Gruneisen and consider an alternative scheme based on Landauer transmission between individual dopants hosting large polarons. We find a scaling relationship between carrier mobility, the electric permittivity and the frequency of transverse optical soft mode in this temperature range. Providing an account of this observation emerges as a challenge to theory.
\end{abstract}

\maketitle

\begin{figure}
\includegraphics [width=0.45\textwidth]{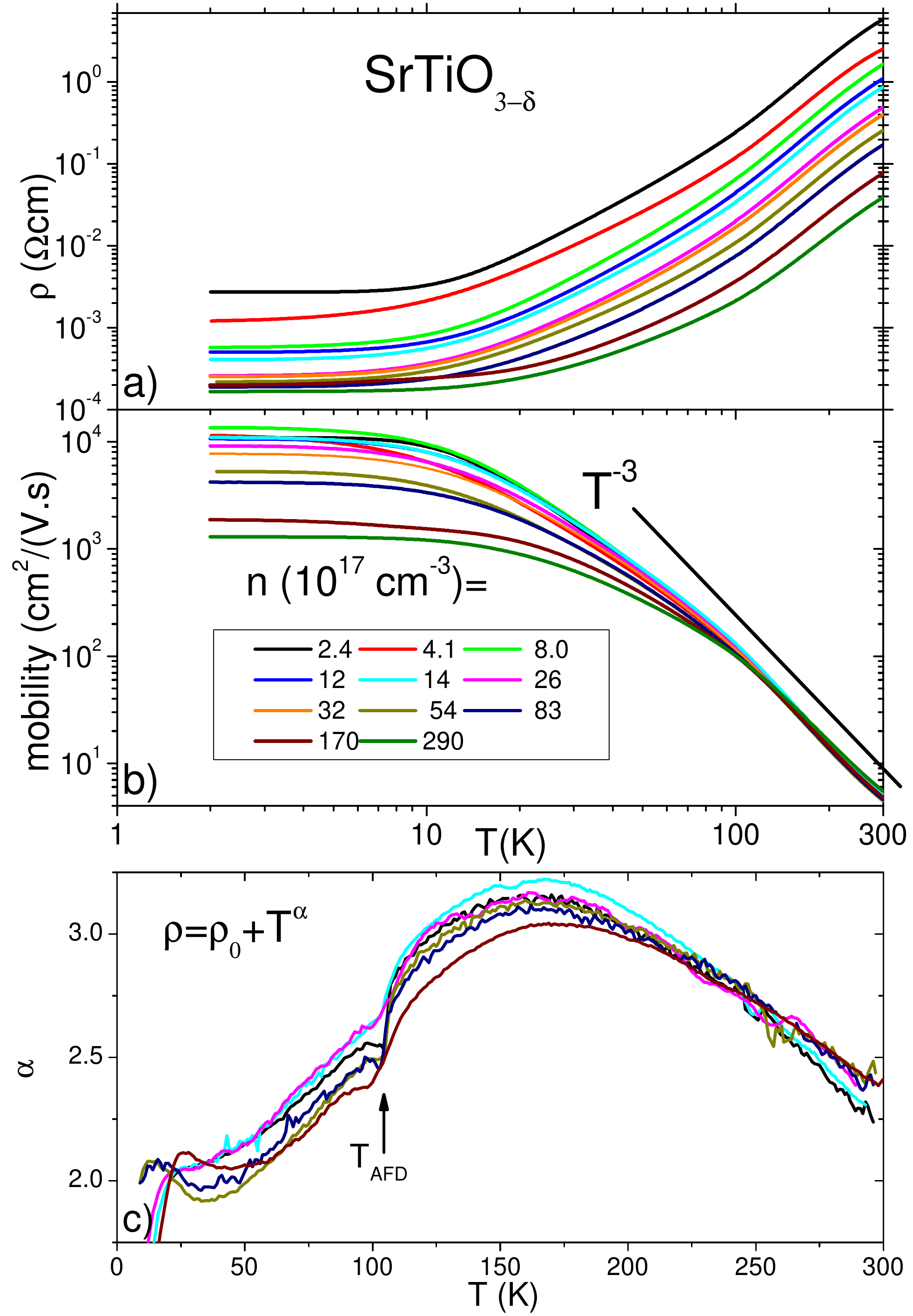}
\caption{\textbf{Resistivity and mobility in doped SrTiO$_3$:} a) Temperature dependence of resistivity in  SrTiO$_{3-\delta}$ as the carrier concentrations is tuned from 10$^{17}$ to 10$^{19}$ cm$^{-3}$. Room-temperature resistivity is several hundred times higher than low-temperature resistivity. b) Hall mobility as a function of temperature. Above 100 K, the mobility does not depend on carrier concentration and is roughly cubic in temperature. c) Assuming that inelastic resistivity follows a power law, i.e. $\rho= \rho_{0}+ AT^{\alpha}$, one can extract the exponent, $\alpha$ by taking a logarithmic derivative: $\alpha=dln(\rho-\rho_{0})/dlnT$, as in the case of cuprates\cite{Cooper:2009}. At low temperature, $\alpha\simeq 2$. Above 50K, it starts a significant shift upward  and  exceeds 3 around 150K, before steadily decreasing afterwards. The antiferrodistortive transition\cite{Tao:2016} is the source of the small anomaly at 105K.}
\label{Fig1}
\end{figure}

\section{Introduction}
The existence of well-defined quasi-particles is taken for granted in the Boltzmann-Drude picture of electronic transport. In this picture, carriers of charge or energy are scattered after traveling a finite distance. The Mott-Ioffe-Regel(MIR) limit is attained when the  mean-free-path of a carrier falls below its Fermi wavelength or the interatomic distance\cite{Hussey:2004}. In most metals, resistivity saturates when this limit is approached. But in ``bad''\cite{Emery:1995} or ``strange''\cite{Sachdev:2015} metals, it continues to increase\cite{Gunnarsson:2003}. It is indeed strange when the mean-free-path persists to fall after attaining its shortest conceivable magnitude and often unexpected behavior is considered bad.

In most cases, bad metals present a linear temperature dependence beyond the MIR limit. Bruin \emph{et al.}\cite{Bruin:2013} recently noticed that the T-linear scattering rate in numerous metals, either ordinary or strange, has a similar magnitude. This observation suggests the relevance of a universal Planck timescale ($\tau_{P}\sim\frac{\hbar}{k_{B}T}$, where $k_{B}$ is the Boltzmann and $\hbar$ the reduced Planck constants) to electronic dissipation and motivated a theoretical attempt to quantify an upper limit to the diffusion constant in incoherent metals\cite{Hartnoll:2014}. During the past years, several theories of charge transport with no need for well-defined quasi-particles were proposed\cite{Mukerjee:2006,Lindner:2010,Andreev:2011,Davison:2014,Principi:2015,Pakhira:2015,Lim:2015,Perep:2016}. More recently, direct measurements of thermal diffusivity\cite{Zhang:2016} have provided additional evidence for quasi-particle-free thermal transport at room temperature in cuprates.

We show here that n-doped SrTiO$_{3}$ is an unnoticed case of strange or bad metallicity. The resistivity of this dilute metal decreases by orders of magnitude upon cooling to cryogenic temperatures\cite{Fredrikse:1964,Tufte:1967,Fredrikse:1967,Wemple:1969,Spinelli:2010}. In the low-temperature limit, resistivity follows a T-square behavior\cite{Vandermarel:2011,Lin:2015}, as expected in a Fermi liquid. Above 100 K, however, the temperature dependence is close to cubic\cite{Tufte:1967,Lin:2015}. In this regime, the mean-free-path of carriers becomes shorter than their wavelength. At room temperature, it falls well below the interatomic distance, the lowest conceivable length scale. Strontium titanate is the strangest known metal according to the criteria for strangeness widely used. Since quasi-particles in this system are not well-defined in a time scale long enough to allow the existence of a scattering time, charge transport cannot be reasonably described by a phonon-scattering picture. This was the approach in previous attempts\cite{Fredrikse:1964,Tufte:1967,Fredrikse:1967,Wemple:1969,Verma:2014, Himmetoglu:2014,Zhou:2016} to understand charge transport. We argue that  metallicity is caused by the temperature dependence of the transmission coefficient along Landauer canals between dopants. Such a picture does not require a scattering time or a mean-free-path. We find an empirical link between mobility and permittivity, which gives a reasonable account of the experimental data. Providing a quantitative account of this observation emerges as a specific challenge to transport theory.

\begin{figure}
\includegraphics [width=0.45\textwidth]{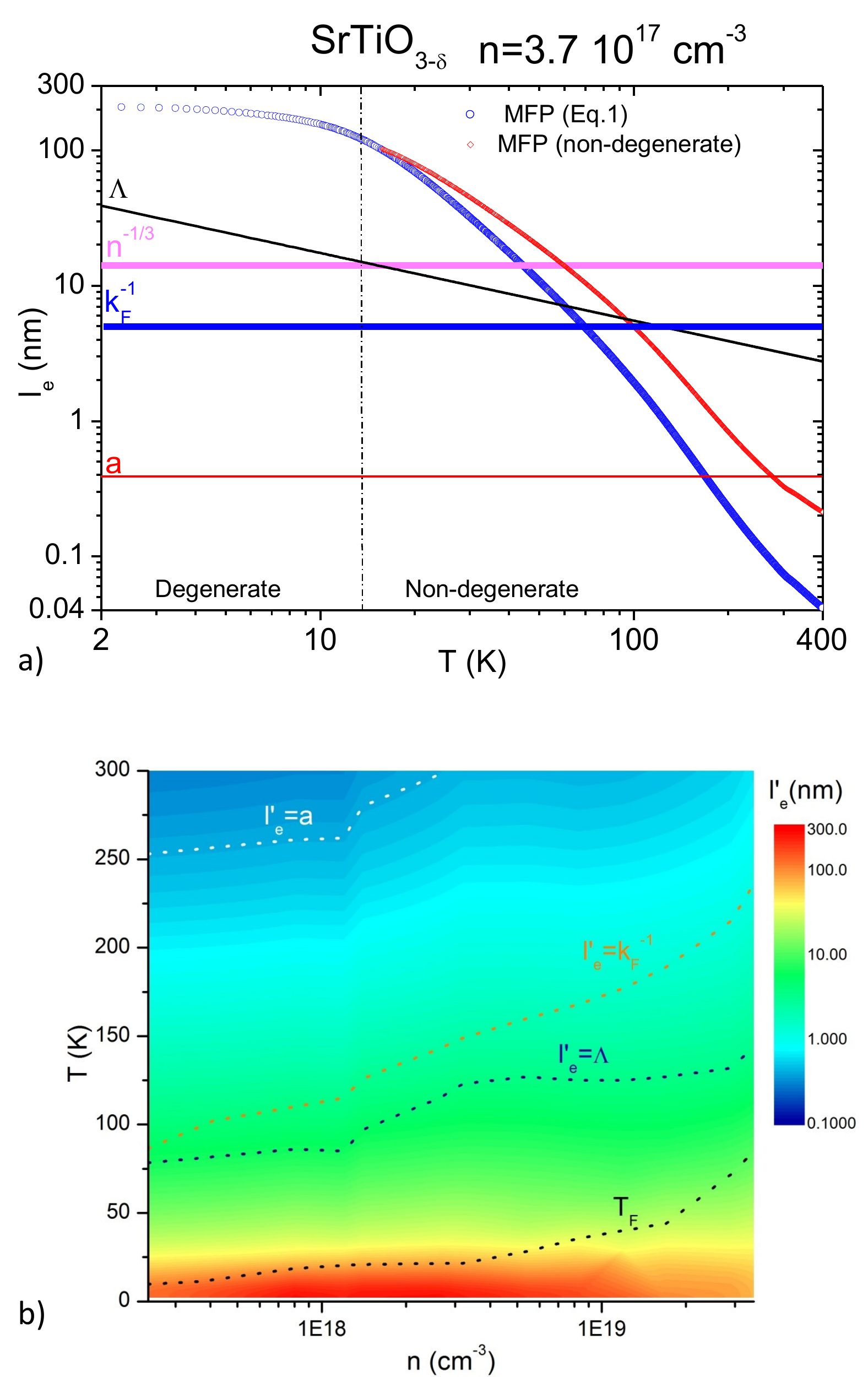}
\caption{\textbf{Breakdown of the Mott-Ioffe-Regel limit:} a) Temperature dependence of the mean-free-path in dilute metallic SrTiO$_{3-\delta}$ sample up to 400 K. Blue circles represent $\ell_{e}$, extracted from resistivity using Eq.1. Red diamonds represent the mean-free-path in the non-degenerate regime: $\ell'_{e}=\ell_{e}\frac{n^{-1/3}}{\Lambda}$. Also shown are the de Broglie thermal wavelength, $\Lambda$, the interelectron distance n$^{-1/3}$, the inverse of the Fermi wavelength ($k_{F}^{-1}$) and the lattice parameter, $a$=0.39 nm. The system remains metallic even at 400 K, in spite of exceeding the MIR limit by all conceivable criteria. b) A color plot of $\ell'_{e}$, extracted from the resistivity data of Fig. 1 in the (T, n) plane. Different crossovers are shown. The MIR limit is exceeded in a high-temperature window narrowing down with increasing concentration.}
\label{Fig2}
\end{figure}

\begin{table*}
\caption{\textbf{How the MIR limit of metallicity is exceeded in different perovskytes-} The table compares the Fermi wave-vector, the resistivity and the mean-free-path at 400 K in four different systems using equation 1. The table lists an underdoped, an optially-doped cuprate and strontium ruthenate and compares them with doped strontium titanate. In  the case of YBa$_{2}$Cu$_{4}$O$_{8}$, we assume that there are four Fermi surface pockets with a k$_F$  corresponding to the square root of the frequency of quantum oscillations. In contrast to other systems, the MIR limit in doped strontium titanate occurs well below room temperature.}
\centering 
\begin{tabular}{|c| c| c |c| c |c |c |} 
  \hline
 Compound & k$_{F}$ (nm$^{-1}$) & $\rho$(400K) (m$\Omega$cm) & $\ell$(400K)(nm)& k$_{F}$$\ell$ (400K) & T$_{k_F\ell=1}$ (K) & Reference\\
  \hline
 La$_{1.72}$Sr$_{0.18}$Cu0$_{4}$ &6.6 & 0.32 & 0.35 & 2.2 & 850 & \cite{Takagi:1992,Gunnarsson:2003}\\
 \hline
  YBa$_{2}$Cu$_{4}$O$_{8}$ & 1.4 & 0.36 & 0.8 & 1.1 & 450 & \cite{Hussey:1997,Bangura:2008}\\
  \hline
  Sr$_{2}$RuO$_{4}$ & 5.6 & 0.2 & 0.38 & 2.1 & 800 & \cite{Tyler:1998}\\
  \hline
  SrTiO$_{3-\delta}$ (n=4 10$^{17}$ cm$^{-3}$)& 0.24 & 5900 & 0.2 & 0.05 & 100 & This work \\
  \hline
\end{tabular}
\label{table:1} 
\end{table*}
\section{Results}
\subsection{Metallicity beyond the MIR limit in strontium titanate}

Fig. 1a shows the temperature dependence of resistivity in SrTiO$_{3-\delta}$ as the carrier concentration is changed from 2.4 10$^{17}$cm$^{-3}$ to 3.5 10$^{19}$cm$^{-3}$. At low temperature ($T\ll100K$), the system displays a T-square resistivity\cite{Vandermarel:2011,Lin:2015} as expected for electron-electron scattering\cite{Pal:2012} in a Fermi liquid. Plotting resistivity as a function of T$^{2}$ (see Fig. 2 in ref.\cite{Lin:2015}), one can see that inelastic resistivity becomes T-square in the low-temperature limit. A wide range of dense Fermi liquids display Kadowaki-Woods scaling\cite{KW} of their T-square resistivity prefactor and their electronic specific heat. Since the latter is proportional to carrier concentration, the case of low-density systems such as doped SrTiO$_{3}$ is different\cite{Hussey:2005}. However, the scaling between the resistivity prefactor and Fermi energy persists. Indeed, the magnitude of this prefactor decreases by several orders of magnitude with changing carrier concentration\cite{Lin:2015}.

Fig. 1b  shows the temperature dependence of the extracted Hall mobility, $\mu=\frac{1}{ne\rho}$. Above 100 K, it displays a temperature dependence close to T$^{-3}$ and barely varies with carrier concentration. Our data is in agreement with the less extensive data reported previously by Tufte and Chapman\cite{Tufte:1967}.  Fig. 1c shows how the exponent of inelastic resistivity evolves as a function of temperature. The total resistivity can be expressed as $\rho=\rho_{0}+AT^{\alpha}$. As seen in Fig. 1c, $\alpha$ is 2 at very low-temperature ($T<50K$) but smoothly rises with warming. One does not expect the T-square resistivity to survive at the Fermi degeneracy temperature. Therefore, the electron-electron scattering origin of the T-square term in the dilute limit has been put into question\cite{Maslov2017,Swift2017}. We note, however, that the smooth variation of $\alpha$, makes the detection of the upper boundary of T-square resistivity impossible. As one can see in the figure, $\alpha$ peaks around 150 K at a value slightly exceeding 3 and then  decreases with increasing temperature.

In the zero temperature limit, the magnitude of electron mobility, set by defect scattering can become remarkably large in doped strontium titanate\cite{Son:2010}.  As a consequence, quantum oscillations are easily detectable in moderate magnetic fields\cite{Lin:2013,Allen:2013,Lin:2014}. The large magnitude of defect-limited mobility at low temperature and its evolution with carrier concentration can be explained in a simple model\cite{Behnia:2015} invoking the long effective Bohr radius of the quantum paraelectric\cite{Muller:1979} parent insulator.

The focus of the present paper is charge transport above 100 K where mobility presents a strong temperature dependence and almost no variation with carrier concentration over two orders of magnitude (Fig. 1b). Early studies\cite{Fredrikse:1964,Tufte:1967,Fredrikse:1967,Wemple:1969} attributed the T$^{-3}$ variation of mobility to the scattering of carriers by phonons. There was a controversy between Fredrikse \emph{et al.}\cite{Fredrikse:1967} (who invoked longitudinal optical phonons) and  Wemple \emph{et al.}\cite{Wemple:1969} (who argued in favor of transverse optical phonons). Neither during this early debate\cite{Fredrikse:1964,Tufte:1967,Fredrikse:1967,Wemple:1969}, nor in more recent theoretical accounts of the temperature dependence of mobility in n-doped SrTiO$_{3}$\cite{Baratoff:1981,Verma:2014, Himmetoglu:2014,Mikheev:2015,Zhou:2016}, the magnitude of the mean-free-path was discussed.

In the Drude-Boltzmann picture, the conductivity, $\sigma$, of a metal depends on fundamental constants and a number of material-related properties. These are either, the scattering time $\tau$, and the effective mass, $m^{*}$, or the Fermi wave-vector, k$_{F}$ and the electron mean-free-path, $\ell_{e}$. Namely:
\begin{equation}\label{1}
\sigma=\frac{ne^{2}\tau}{m^{*}}=\frac{1}{3\pi^{2}}\frac{e^{2}}{\hbar}k_{F}^{2}\ell_{e}
\end{equation}

The  Fermi surface of n-doped SrTiO$_{3}$ in the dilute limit is known. It is located at the center of the Brillouin zone, because a band originating from Ti atoms' t$_{2g}$ orbitals has a minimum at the $\Gamma-$point\cite{Vandermarel:2011,Allen:2013,Tao:2016}. The threefold degeneracy of this band is lifted by tetragonal distortion and spin-orbit coupling.  As the system is doped, the three bands are filled successively and three concentric Fermi surfaces will emerge and grow in size one after the other. This is a picture based on band calculations\cite{Vandermarel:2011} and confirmed by experimental studies of quantum oscillations\cite{Lin:2013,Allen:2013,Lin:2014}. Below 10$^{18}$cm$^{-3}$, the first critical doping for the emergence of secondary pockets, there is a single Fermi surface with moderate (1.6) anisotropy\cite{Allen:2013,Lin:2014,Tao:2016}. Therefore, k$_F$ can be quantified in a straightforward manner. Fig. 2a shows the temperature dependence of the mean-free-path extracted from  Eq. 1 in a dilute sample whose resistivity was measured up to 400 K. The carrier density is the one given by the Hall coefficient. As previously reported in ref. \cite{Tufte:1967} , the Hall coefficient does not display any notable temperature dependence. Moreover, the carrier density according to the magnitude of the Hall coefficient matches the Fermi surface volume according to the frequency of quantum oscillations\cite{Lin:2014}. Therefore, there is little ambiguity regarding carrier concentration.

As seen in Fig. 2a, the mean-free-path becomes shorter than the inverse of the Fermi wave-vector k$_{F}^{-1}$ and then the lattice parameter $a$. Now, at this carrier concentration, with an effective mass of m$^{*}$=1.8m$_{e}$, the Fermion degeneracy temperature  is as low as 15 K\cite{Lin:2013}. As one sees in the figure, this is the temperature at which de Broglie thermal wavelength, $\Lambda=\frac{h}{\sqrt{2\pi m^{*}k_{B}T}}$ becomes comparable to the interelectron distance $n^{-1/3}$. Above this temperature, the electrons become non-degenerate and follow a Maxwell-Boltzmann distribution. As a consequence and thanks to thermal excitation, the typical carrier momentum is larger than $\hbar$k$_{F}$.  In other words, the typical velocity is not the Fermi velocity, but the thermal velocity, $v_{T}= \sqrt{\frac{2k_{B}T}{m^{*}}}$. Let us define the mean-free-path of electrons with thermal velocity as $\ell'_{e}=\tau v_{T}=\ell_{e}n^{-1/3}/\Lambda$. As seen in the figure,  because of $\ell'_{e}>\ell_{e}$, the crossover temperatures shift upward. However, the picture remains qualitatively the same. In the non-degenerate regime, the relevant  wavelength is $\Lambda$ and not the Fermi wavelength. As seen in the figure, the mean-free-path becomes shorter than $\Lambda$ too.

A room-temperature resistivity of 2.4 m$\Omega$.cm implies a scattering time of $\tau \simeq 6fs$. This is shorter than the Planck time\cite{Bruin:2013} ($\tau_{P}=\frac{\hbar}{k_{B}T}$) at room temperature, $\tau_{P}(300K)=25.3 fs $. This is the typical time it takes for a carrier with a thermal velocity, v$_{T}$, to move as far as its de Broglie thermal wavelength ( $\frac{\Lambda}{v_{T}}=1.77\tau_{P}$). By all accounts, room-temperature quasi-particles in dilute metallic strontium titanate do not live long enough to qualify for a scattering-based picture.

A color plot of the mean-free-path in the (T, n) plane produced from the data of Fig.1 and Eq. 1 is presented in Fig. 2b. The MIR limit is exceeded in  a high-temperature window, widening with the decrease in carrier concentration.  The hitherto unnoticed specificity of SrTiO$_{3}$ is that it becomes a strange metal well below room temperature. The data in Table 1 gives an idea of the ``strangeness'' or the ``badness'' of SrTiO$_{3}$ compared to two other perovskytes. As seen in the table, at 400 K, k$_{F}\ell$ in this system is much lower than in cuprates or in Sr$_{2}$RuO$_{4}$. Contrary to the two other systems, the passage beyond the MIR limit occurs well below room temperature implying the necessity of a transport picture with no reference to well-defined quasi-particles. This is the first principal message of this paper.

\begin{figure}
\includegraphics [width=0.42\textwidth]{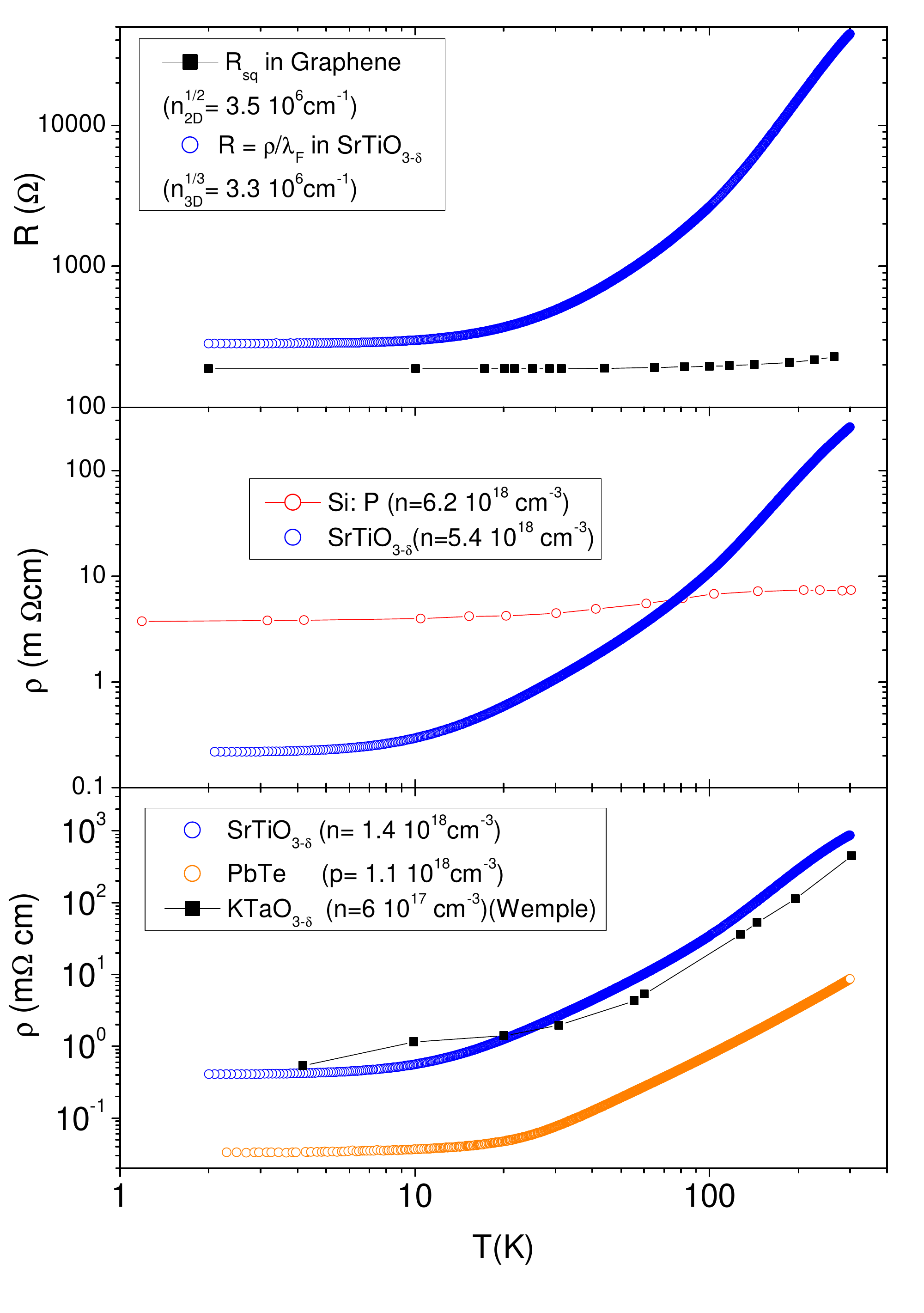}
\caption{\textbf{Comparison with other dilute metallic systems:} a) R$_{\Box}$ of a one Fermi-wavelength-thick sheet of  SrTiO$_{3-\delta}$ compared with graphene\cite{Efetov:2010} with comparable areal carrier density. b) Temperature-dependence of resistivity in SrTiO$_{3-\delta}$ and in metallic silicon\cite{Yamanouchi:1967} with comparable volume carrier concentration. c) Temperature-dependence of resistivity in SrTiO$_{3-\delta}$,  metallic PbTe and metallic KTaO$_{3-\delta}$\cite{Wemple:1965} in similar range of carrier concentration. In silicon and graphene, the change in resistance caused by phonon scattering of electrons is a modest part of the total resistance. In  all other systems, which are close to a ferroelectric instability, resistivity changes by several orders of magnitude.}
\label{Fig3}
\end{figure}

\subsection{Comparison with graphene and metallic silicon}
The temperature dependence of resistivity above 100 K and the magnitude of inelastic resistivity are very different from dilute metallic systems subject to phonon scattering. Let us compare strontium titanate with two exemplary systems. A recent study of charge transport in graphene\cite{Efetov:2010} has shown the relevance of the Bloch-Gr\"{u}neisen picture of electron-phonon scattering in our temperature window of interest. In this picture, above a characteristic temperature (usually, but not always, the Debye temperature) resistivity displays a linear temperature dependence. This happens because all phonon modes capable of scattering electrons are thermally populated and the phase space for scattering is proportional to the thermal width near the Fermi level. Far below the characteristic temperature, the resistivity is expected to display a much stronger T$^{5}$ (in 3D) or T$^{4}$ (in 2D) temperature dependence because the typical wave-vector of the thermally excited phonons becomes smaller with decreasing temperature. As a consequence, the capacity of scattering phonons to decay the momentum flow decays rapidly with temperature. Efetov and Kim\cite{Efetov:2010} showed that this picture gives a very satisfactory account of inelastic resistivity in graphene as the carrier density is tuned from 1.36 to 10.8 10$^{13}$cm$^{-2}$. Let us note that in gated graphene, inelastic (that is, temperature-dependent) resistivity is  far from dominating the total resistivity. This is seen in Fig. 3a, which compares the temperature-dependence of resistivity in n-doped SrTiO$_{3}$ and graphene at an identical interelectron distance. As one can see in the figure, resistivity changes by 200 in SrTiO$_{3}$ and by 1.6 in  graphene. Charge conductivity is higher in graphene by a factor of two at low temperature and by two orders of magnitude at room temperature. A comparison with the case of metallic silicon\cite{Yamanouchi:1967} reveals a similar discrepancy. As one can see in Fig. 3b, cooling  phosphorous-doped silicon leads to a modest change in resistivity, much less than what is seen in doped SrTiO$_{3}$. As a consequence, while at low-temperature, strontium titanate has a significantly higher mobility, the opposite is true at room temperature. In contrast to strontium titanate, in both graphene and silicon, one can use Eq. 1 one to extract a meaningful mean-free-path, which increases by a factor of two or less as the system is cooled down from room temperature to helium liquefaction temperature.

It is also instructive to compare SrTiO$_{3}$ with two other dilute metals close to a ferroelectric instability. As one can see in Fig.3c, both PbTe and KTaO$_{3}$ show comparably large changes in their resistivity upon cooling. Reports on KTaO$_{3}$\cite{Wemple:1965,Wemple:1969} and on lead chalcogenide salts (PbTe, PbSe and PbS)\cite{Petritz:1955,Allagier:1958} indicate that high-temperature mobility does not vary with carrier concentration, but shows a strong temperature dependence ($\mu \propto T^{-\lambda}$, with $\lambda\simeq2.5-3$). Both these features are similar to what we see in strontium titanate. Only in strontium titanate, however, the magnitude of the room temperature mobility is low enough to make it a clearly strange metal below room temperature.

To sum up, in silicon and graphene, the temperature-induced change in resistivity implies a modest change in mean-free-path and follows a Bloch-Gruneisen behavior (a slow T-linear behavior at high temperature and a fast T$^{4}$ or T$^{5}$ behavior at low temperature). On the other hand, doped SrTiO$_{3}$ and other metals close to a ferroelectric instability do not fit within this picture. The apparent mean-free-path of carriers changes by orders of magnitude. It becomes very short at high temperature, and in the case of strontium titanate, too short to be physically meaningful.
\section{Discussion}
\subsection{Conduction as transmission in a doped polar semiconductor}
\begin{figure}
\includegraphics [width=0.45\textwidth]{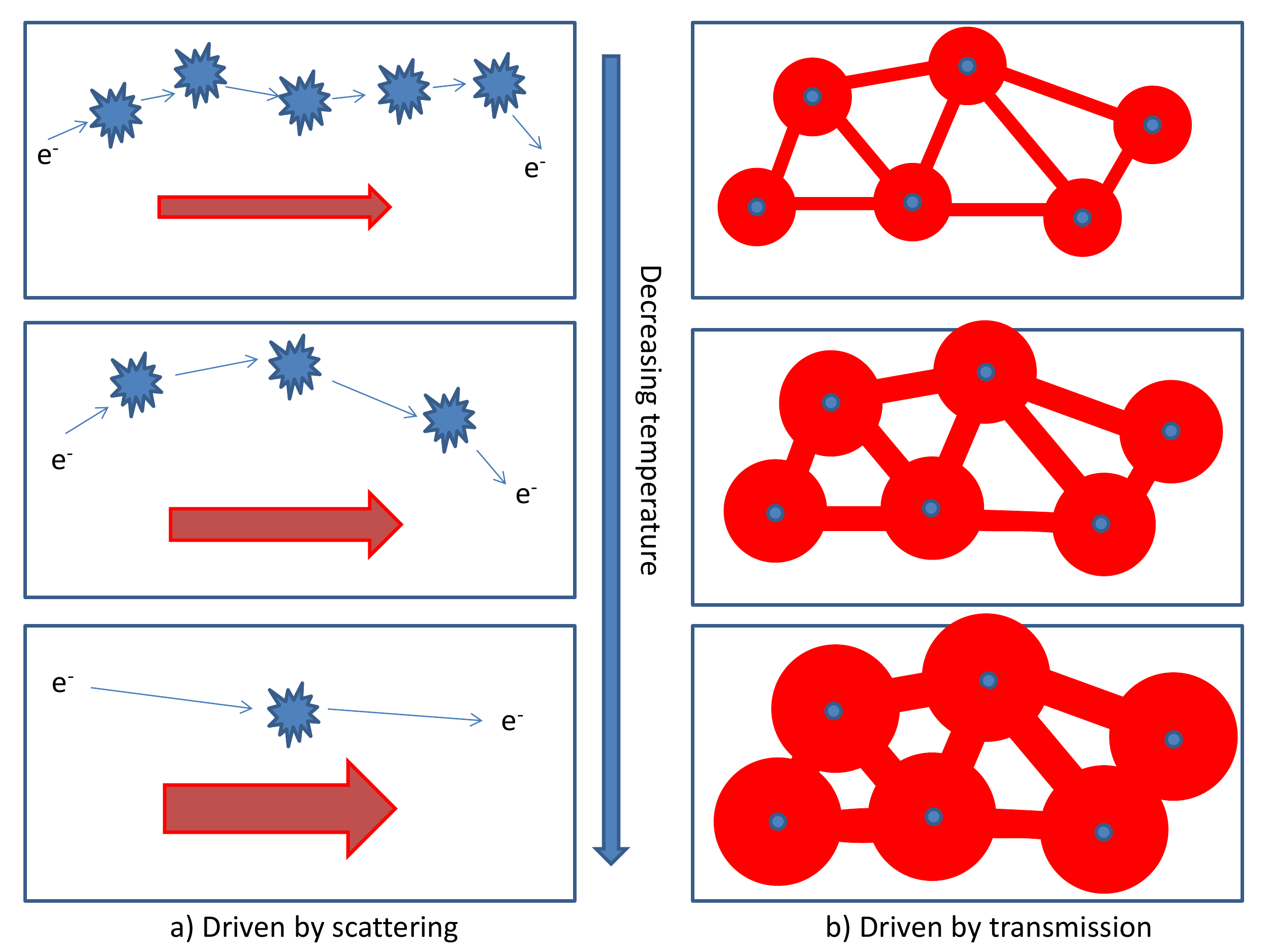}
\caption{\textbf{Two pictures of temperature-dependent conductivity:} a) In a Drude-Boltzmann picture of transport, cooling the system leads to a rarefication of scattering events. Conductivity increases because the time interval (and the spatial distance) between two successive scattering events increases with cooling. Such a picture requires  a well-defined quasi-particle across the distance in time and in space separating two scattering events. b) In a Landauer picture of transport, each dopant (each oxygen vacancy here) is a reservoir connected to other adjacent reservoirs by conducting channels. Conductivity increases upon cooling because the transmission coefficient enhances. Such a picture does not require well-defined quasi-particles. }
\label{Fig4}
\end{figure}

Fig. 4 sketches two alternative descriptions of charge conduction. In the semi-classic picture of metallic charge transport (Fig. 4a), conductivity increases with decreasing temperature, because the scattering events relaxing the momentum of a charge-carrying quasi-particle become less frequent with decreasing temperature. Now, consider an alternative picture (Fig. 4b),  where each dopant is hosting a polaron, `an electron trapped by digging its own hole'\cite{Mott:1940} in a polar semiconductor. The polaron picture has been frequently invoked in the case of strontium titanate\cite{Eagles:1965,Mechelen:2008}.

In this second picture, conductivity is set by transmission among dopant sites and there is no need for well-defined quasi-particles. Can conductivity increase with decreasing temperature in such a picture and generate a metallic behavior? This can happen if cooling increases the transmission probability between adjacent dopant sites. In the case of strontium titanate, static permittivity increases with cooling\cite{Muller:1979}. Therefore, the lower the temperature, the shallower the donor potential. As a consequence, polarons become less bound with decreasing temperature and conductivity is metallic.

\begin{figure}
\includegraphics [width=0.42\textwidth]{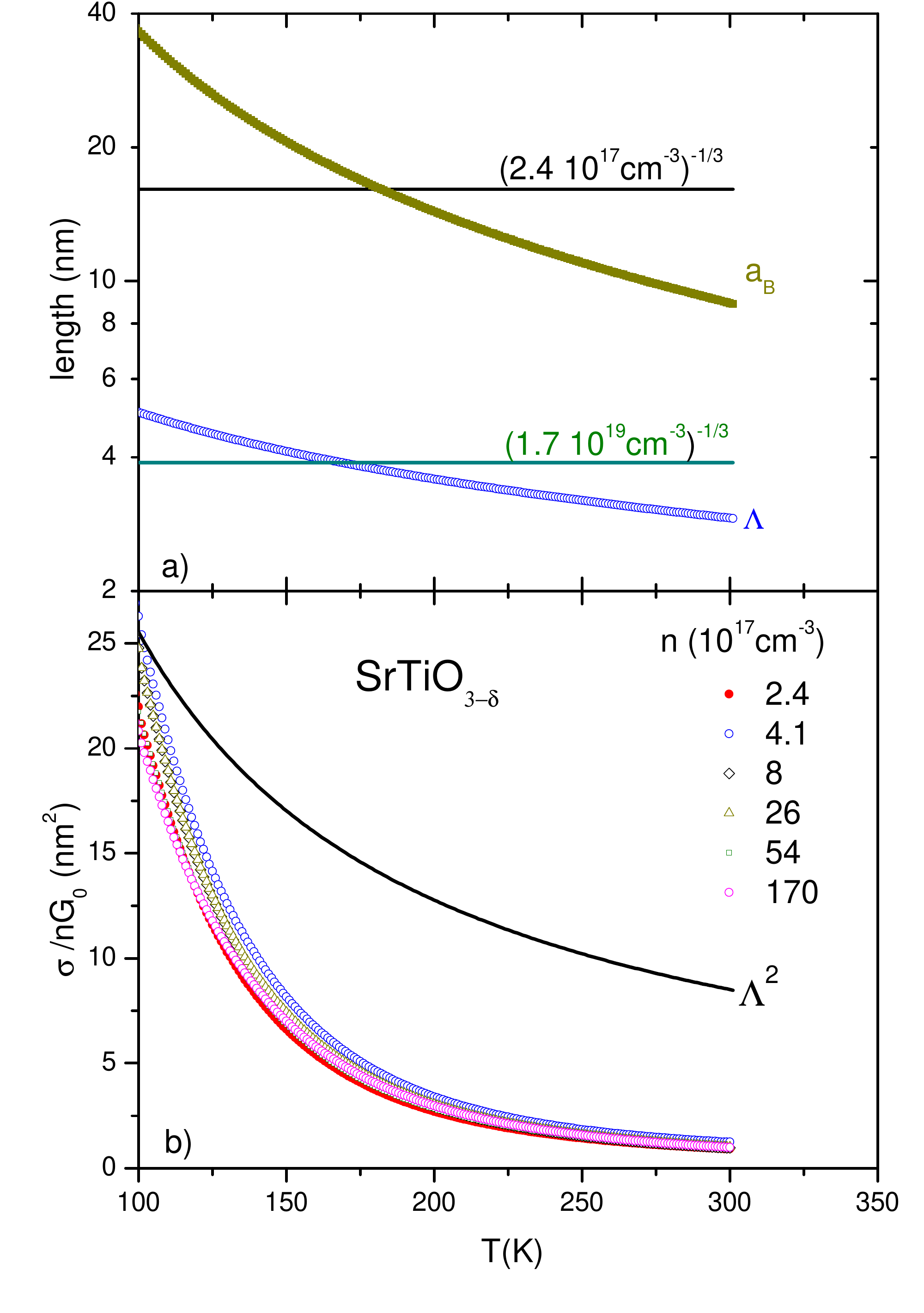}
\caption{\textbf{Length scales and transmission} a) Temperature dependence of different length scales at two different carrier densities. The de Broglie thermal wavelength, $\Lambda$, was calculated assuming an effective mass of m$^{*}$=1.8m$_{e}$, found by quantum oscillations at low temperature. The effective Bohr radius was calculated using this mass and the permittivity data reported in \cite{Muller:1979}.  b) Transmission coefficient extracted from resistivity data for different carrier concentrations compared to  $\Lambda^{2}$.}
\label{Fig5}
\end{figure}

Let us now consider different length scales of the system. Fig. 5a shows them in our temperature range of interest where mobility becomes independent of carrier concentration. When carrier density changes from $2.4\times 10^{17}cm^{-3}$ to $1.7\times 10^{19}cm^{-3}$, the distance between electrons varies between 16 to 4 nm. Taking the low-temperature effective mass of m$^{*}$=1.8m$_{e}$ seen by quantum oscillations\cite{Lin:2013}, one finds that the de Broglie thermal wave-length is $\Lambda\simeq 3-6 nm$ in this temperature range. Thus, the system is mostly non-degenerate. Electric permittivity, $\epsilon$, is large ($\epsilon(300K) \simeq300)$\cite{Muller:1979, Lowndes:1973} and increases steadily with cooling. As a consequence, the effective Bohr radius (a$^{*}_{B}=\frac{4\pi\epsilon\hbar^{2}}{m^{*}e^{2}}$) varies between 8 and 40 nm.  This puts the system above the Mott criterion for metallicity (n$^{1/3}$ a$^{*}_{B}> 0.25$\cite{Edwards:1978}), which, one shall not forget, is relevant to degenerate electrons\cite{Mott:1990}. These length scales are in the same range of magnitude. They are all longer than the lattice parameter (0.39nm). The question is how to picture conduction as transmission\cite{Imry:1999} along a random network of donors.

Since mobility is found to be independent of carrier concentration above 100K, we can express conductivity as:

\begin{equation}\label{2}
\sigma=\rho^{-1}=G_{0}n <T>
\end{equation}

Here, $G_{0}=\frac{2e^{2}}{h}$, is the quantum of conductance and $<T>$ represents the average transmission between a dopant site and its immediate neighbors. Note that in three dimensions, $<T>$ is proportional to mobility  ($\mu=2\frac{e}{h}<T>$).

The temperature dependence of $<T>$,  extracted from the experimental data using Eq. 2 is shown in Fig. 5b. As seen in the figure, a hundred-fold enhancement in carrier concentration does not affect $<T>$. One would have expected that the Landauer transmission between one site and its more or less distant neighbors would depend on the length of the trajectory and the details of the structural landscape. It is therefore surprising that the averaged transmission coefficient, $<T>$, is insensitive to such a large change in the average interdopant distance.

Fig. 5b also compares the magnitude of $<T>$  with $\Lambda^{2}$. the latter represents the areal uncertainty in the spatial localisation of a non-degenerate electron in three dimensions. As seen in the figure, $<T>$ decreases faster than $\Lambda^{2}$. There is little experimental information regarding the effective mass in this temperature range. A larger temperature-dependent effective mass would make $\Lambda^{2}$ smaller and its decrease faster. Thus, one can bridge the gap between $<T>$ and  $\Lambda^{2}$ with a hypothetical temperature-dependent polaronic  mass. However, there is no experimental evidence for this. At low temperatures, quantum oscillations\cite{Lin:2014, Allen:2013} resolve a modest mass enhancement($\leq4$). This is also the case of infrared conductivity\cite{Mechelen:2008}. Thus, we cannot simply picture conductivity as ballistic charge transport along Landauer wires. Let us now consider a diffusive picture.

\subsection{Mobility, diffusivity and their link to electric permittivity}

In a non-degenerate semiconductor, mobility and diffusion constant, $D$, of carriers are linked through Einstein's equation\cite{Ashcroft:1976}:

 \begin{equation}\label{3}
D=\frac{\mu k_{B}T}{e}
\end{equation}

The temperature dependence of $D$  extracted from the mobility data of the sample with a carrier concentration of 2.4 10$^{17}$ cm$^{-3}$ is shown in Fig. 6a. Note that since mobility does not show any detectable variation with  carrier concentration, this curve is representative of $D$ for carrier concentrations in the range of 10$^{17}$-10$^{19}$cm$^{-3}$.  The magnitude of $D$ obtained in this way is remarkably low, well below 1cm$^{2}$/s. For comparison,  the diffusion constant of electrons in silicon rises from 35 cm$^{2}$/s at room temperature to $\sim$100 cm$^{2}$/s at 100 K\cite{Brunetti:1981}.

Hartnoll recently  proposed a lower bound to diffusion constant in incoherent metals\cite{Hartnoll:2014}:

\begin{equation}\label{4}
D_{H}=\frac{\hbar v_{F}^{2}}{k_{B}T}
\end{equation}

As one can see in Fig. 6a, the experimental $D$ decreases faster than T$^{-1}$. Hartnoll boundary was obtained for degenerate systems. Ours is non-degenerate and the typical velocity is thermal velocity and not v$_{F}$. The Fermi velocity changes drastically with carrier density, but the experimental $D$ does not. As defined in Eq. 4 $, D_{H}$,  does  not appear to be directly relevant to our observation.

\begin{figure}
\includegraphics [width=0.42\textwidth]{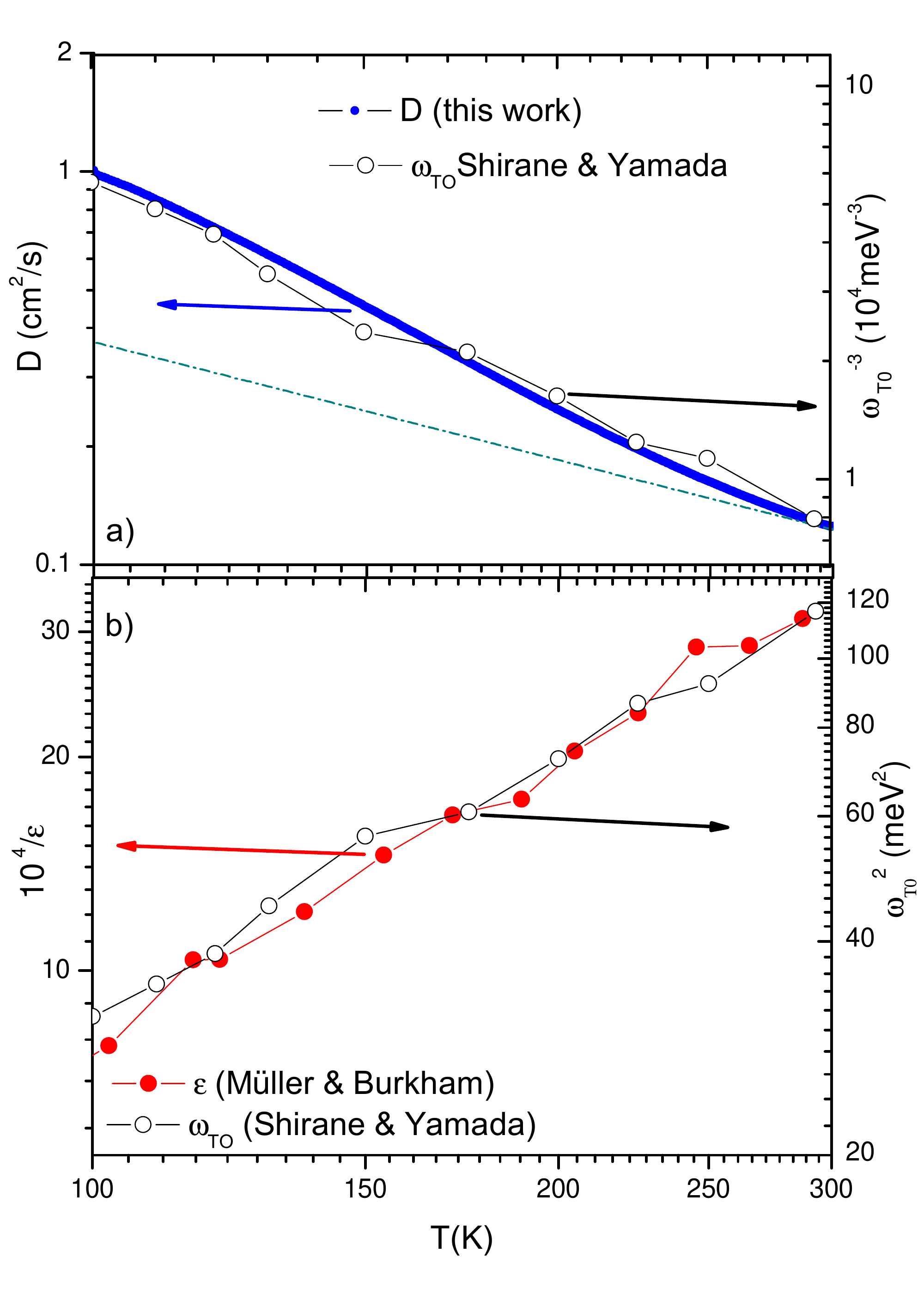}
\caption{\textbf{Diffusivity, ferroelectric soft mode and permittivity} a) Blue solid circles represent the diffusion constant extracted from mobility. The dashed line represents a T$^{-1}$ temperature dependence. Open circles represent the soft mode frequency, $\omega_{TO}^{-3}$ according to ref.\cite{Fleury:1968}.  The two quantities scale with each other in this temperature range ($D\propto\omega_{TO}^{-3}$). b) The inverse of electric permittivity, $\epsilon^{-1}$\cite{Muller:1979} and $\omega_{TO}^{2}$ in the same temperature range. The scaling is $\epsilon^{-1}\propto\omega_{TO}^{2}$.}
\label{Fig5}
\end{figure}

An obvious piece of the puzzle of charge transport in strontium titanate is the fact that the insulating  parent is a quantum paraelectric with a strongly temperature-dependent electric permittivity\cite{Lowndes:1973,Muller:1979}. We have found that in our temperature range of interest $D$ is proportional to $\epsilon^{1.5} $.  This empirical observation is the second principal message of this paper.

The scaling between static electric permittivity, $\epsilon_{0}$, and the frequency of the Transverse Optic (TO) soft mode\cite{Fleury:1968,Yamada:1969}, $\omega_{TO}$ in strontium titanate is well-known. Yamada and Shirane\cite{Yamada:1969} showed that $\omega_{TO}=194.4\sqrt{\epsilon^{-1}}$ within experimental resolution. In a polar crystal, there is  a theoretical link between the ratio of static, $\epsilon_{0}$, to high-frequency, $\epsilon_{\infty}$, permittivity and the ratio of longitudinal, $\omega_{L}$, to transverse, $\omega_{T}$, phonon frequencies. This is the expression first derived by Lyddane, Sachs, and Teller\cite{Lyddane:1941,Cohran:1962}:

\begin{equation}\label{5}
\frac{\epsilon_{0}}{\epsilon_{\infty}}=(\frac{\omega_{L}}{\omega_{T}})^2
\end{equation}

When the transverse mode is soft, the system begins to show a large static polarizability. As one can see in Fig. 6b, the available experimental data confirm the scaling between these two distinct measurable quantities. A theory assuming an anisotropy in the polarizability of the oxygen ion\cite{Migoni:1976}, the starting point of the polarization theory of ferroelectricity\cite{Bilz:1987}, explains the temperature dependence of $\omega_{TO}$ in strontium titanate quantitatively. The soft mode and the large permittivity are properties of the insulating parent. However, they are known to survive in presence of dilute metallicity.  Indeed, neutron scattering studies have explored the way the soft mode responds to the introduction of mobile electrons\cite{Bauerle,Bussmann:1981} and found negligible shift in frequency for carrier densities below 7.8 10$^{19}$cm$^{-3}$.

As one can see in Fig. 6a, comparing our data with those reported by Yamada and Shirane\cite{Yamada:1969}, one finds that:
 \begin{equation}\label{7}
\mu k_{B}T \propto \omega_{T0}^{-3}\propto \epsilon/\omega_{T0}
\end{equation}

An explanation for this empirical link between carrier mobility and lattice properties is beyond the scope of this paper and emerges as a challenge to theory.

\subsection{Concluding remarks}
The persistence of T-square resistivity in dilute metallic strontium titanate is puzzling\cite{Lin:2015}. However, its context is familiar. Electrons are degenerate and k$_{F}\ell >>1$. We now see that this low-temperature puzzle is only the tip of an iceberg. Above 100 K, the application of the Drude picture of conductivity yields a paradox: a carrier mean-free-path shorter than all conceivable length scales. Both the magnitude and the functional behavior of the temperature-induced change in resistivity are very different from what is expected according to the Bloch-Gr\"{u}neisen law and what is seen in silicon and graphene. The aim of this paper is to attract theoretical attention to a hitherto unnoticed case of strange metallicity. It remains to be seen if electron viscosity plays any role in diffusive transport. On the experimental side, extending the available data to higher temperatures and lower doping concentration would map the boundaries of metallicity and the MIR limit in this system as well as other polar semiconductors.

It is a pleasure to thank Sean Hartnoll and Dmitrii Maslov for stimulating discussions. This work has been supported  by an \emph{Ile de France} regional grant, by Fonds ESPCI-Paris and by JEIP- Coll\`{e}ge de France and funded in part by a QuantEmX grant from ICAM and the Gordon and Betty Moore Foundation through Grant GBMF5305 to KB.

\end{document}